# Efficient Even Distribution of Power Consumption in Wireless Sensor Networks

Ioan Raicu
Department of Computer Science
Purdue University
1398 Computer Science Building
West Lafayette, IN 47907, USA
iraicu@cs.purdue.edu

**ABSTRACT**

*One of the limitations of wireless sensor nodes is their inherent limited energy resource. Besides maximizing the lifetime of the sensor node, it is preferable to distribute the energy dissipated throughout the wireless sensor network in order to minimize maintenance and maximize overall system performance. We investigate a new routing algorithm that uses diffusion in order to achieve relatively even power dissipation throughout a wireless sensor network by making good local decisions. We leverage from concepts of peer-to-peer networks in which the system acts completely decentralized and all nodes in the network are equal peers. Our algorithm utilizes the node load, power levels, and spatial information in order to make the optimal routing decision. According to our preliminary experimental results, our proposed algorithm performs well according to its goals.*

## 1.0     INTRODUCTION

Through our work, we hope to address the limitations of wireless sensor networks such as: limited energy resources, varying energy consumption based on location, high cost of transmission, and limited processing capabilities. Besides maximizing the lifetime of the sensor nodes, it is preferable to distribute the energy dissipated throughout the wireless sensor network in order to minimize maintenance and maximize overall system performance.

Any communication protocol that involves synchronization between peer nodes incurs some overhead of setting up the communication. Obviously, each node could make the most informed decision regarding its communication options if they had complete knowledge of the entire network topology and power levels of all the nodes in the network. This should yield the best performance if the synchronization messages are not taken into account. However, since all the nodes would always have to know everything, it should be clear that there will be many more synchronization messages than data messages, and therefore ideal case algorithms are not feasible in a system where communication is very expensive.

The usual topology of wireless sensor networks involves having many network nodes dispersed throughout a specific physical area. There is usually no specific architecture or hierarchy in place and therefore, the wireless sensor networks are considered as ad hoc networks. That does not mean that a dynamic organization is not allowed, however it just cannot be predefined in advance.

An ad hoc wireless sensor network is an autonomous system of sensor nodes in which all nodes act as routers connected by wireless links. Although ad hoc networks usually imply that nodes are free to move randomly and organize themselves arbitrarily, in this paper we considered only ad hoc networks with fixed node positions.

On the other hand, wireless links are not very reliable and nodes might stop operating at arbitrary points within the system's life; therefore, the routing protocol utilized must be able to handle arbitrary failure of nodes throughout the network. Such a network may operate in a standalone fashion, or it may be connected to other networks, such as the larger Internet.

Eventually, the data retrieved by the sensors must be propagated back to a central location, where further processing must be done in order to analyze the data and extract meaningful information from the large amounts of data. With plentiful available power, each node could theoretically communicate with the base station directly, however due to a limited power supply, spatial reuse of wireless bandwidth, and the nature of radio communication cost which is a function of the distance transmitted squared, it is ideal to send information in several smaller distance wise steps rather than one transmission over a long communication distance [1,2].

Some typical applications for wireless sensors include the collection of massive amounts of sensor data. They can facilitate communication with the user's environment, such as beacons or sources of information throughout the physical space [5]. Some other typical examples are environmental sensors, automotive sensors, highway monitoring sensors, and biomedical sensors.

In our simulation, we use a data collection problem in which the system is driven by rounds of communication, and each sensor node has a packet to send to the distant base station. The diffusion algorithm is based on location, power levels, and load on the node, and its goal is to distribute the power consumption throughout the network so that the majority of the nodes consume their power supply at relatively the same rate regardless of physical location. This leads to better maintainability of the system because of better predictability, and maximizing the overall system performance by allowing the network to function at 100% capacity throughout most of its lifetime instead of having a steadily decreasing node population.

The major issues concerning wireless sensor networks are power management, longevity, functionality, sensor data fusion, robustness, and fault tolerance. Power management deals with extending battery life and reducing power usage while longevity concerns the co-ordination of sensor activities and optimizations in the communication protocol. The functionality deals with future targeted applications. Data fusion encompasses the combining of sensor data, or perhaps data compression. Obviously, robustness and fault tolerance deals with failing nodes and the erroneous characteristic of the wireless communication medium.

Some progress has been made toward overcoming some of these major issues. For example, LEACH (Low-Energy Adap-tive Clustering Hierarchy) [3] and PEGASIS (Power-Efficient GAthering in Sensor Information Systems) [4] have very similar goals compared to what we are proposing. Their biggest shortcomings deal with unrealistic assumptions and algorithms that do not exploit the spatial information regarding the network. It is also worthwhile to mention a routing algorithm, namely directed diffusion [9], which has a similar name but is quite different than our proposed diffusion based algorithm.

The rest of the paper is organized as follows. Section 2 covers our evaluation test-bed; we explain our own realistic assumptions based on real world sensors, the sensors hardware on which we based our assumptions, and the distribution of the sensor nodes. Section 3 covers the description of the proposed diffusion based routing algorithm. We finally conclude with Section 4 in which we talk about future work.

## 2.0 SIMULATION TEST-BED AND ASSUMPTIONS

### 2.1 Software/Hardware Architecture

Our simulation is based on real world wireless sensors, specifically the Rene RF motes designed at University of California, Berkeley (UCB) [6]. We decided to base our work on these sensors purely because they offer a good architecture to validate the findings of this paper. An important aspect of our research is that the details of power consumption or the energy capacity of the battery supply for each wireless sensor is almost irrelevant since the results between the various algorithms should remain relative to each other no matter what the energy related constants are, however these constants could effect the performance gap between various routing algorithms.

### 2.2 Simulation Constants

According to Table 1, we establish the constants that drive the simulation results.

| Description | Value |
|---|---|
| Link bandwidth between peer nodes | 10 Kbit/s |
| 1 message every XXX seconds | 10 second |
| Total size of the packet | 50 bytes |
| Transmission amplification cost for radio transmitter | 1.8 micro joule/bit/m$^2$ |
| Transmission cost for running the radio circuitry per bits processed | 2.51789 micro joule/bit |
| Reception cost for running the radio circuitry per bit processed | 2.02866 micro joule/bit |
| The cost for a mote to be in its idle state; it is a function related to time | 1000 micro joule/second |
| The initial energy each mote is given. | 15390 joule |
| Area of node distribution | 100x100 meter |
| The number of nodes in the network | 100 nodes |

Table 1: Constants utilized throughout the simulation

### 2.3 Node Distribution

In this paper, we established the nodes' positions according to a normal distribution with the mean and the standard deviation both equal to 50. The base station positioned at coordinates (0,0) on the x-axis and y-axis respectively and an area defined by 100X100 meters.

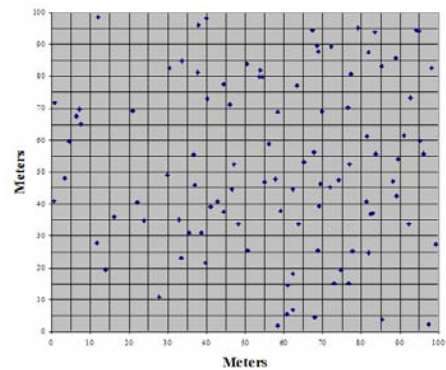

Figure 1: Sample node distribution using an area of 100 meters by 100 meters and having 100 nodes.

## 3.0 DIFFUSION BASED ROUTING ALGORITHM

The most naïve approach would be for all nodes to directly communicate with the base station, regardless of how far the respective node is from the base station. It should be evident that this model will not be suitable for our goals since nodes far away from the base station would deplete their power supply much quicker than nodes that are close to the base station.

A naïve diffusion based routing algorithm based solely on spatial information (eg. distance derived from radio signal strength). Each node would make a list of all available node neighbors and their preferences based on distance. Once this has been established, each node will send to their corresponding neighbor until the receiver's power is depleted, at which time the sender will choose a different neighbor to communicate with. This has the drawback of nodes that are close to the base station having very short life spans since their primary task is to route messages for the entire network, and hence they will die first. Similarly, nodes that are furthest away from the base station will live the longest since they merely communicate with a close neighbor and do not have to relay messages for anyone.

By using location, power levels, and node loads, we can make the local greedy decisions more effective therefore achieve efficient, even power dissipation distribution across all nodes. Each node will act as a relay as each node forwards messages of its peer nodes towards the base station. Each node will locally decide the best neighbors to communicate with using radio signal strength, battery power, and load at the neighboring nodes. The main neighbor selection (based on radio signal strength) would happen at the startup of the system. Afterwards, when a series of events occur, there will be synchronization messages between each sender and receiver that will allow the senders to update their neighbor information with their battery power and load. There will be no keep alive messages since we will assume that nodes do not arbitrarily fail and that they only fail due to inadequate power levels. If we want to address arbitrary failures of nodes, which might be a more realistic assumption, we should use some form of alive messages, whether implicit or explicit.

The synchronization messages that would advise a sender to seek another neighbor to transmit to would occur when certain conditions would be met. The first condition would be if the receiver's power levels are lower than that of the sender. Notice that this can be calculated easily without any extra messages being transmitted since the sender can include its power level information along with the data. The second condition would be if the receiver's power levels are drained below a certain threshold in which the receiver does not want to service any more requests other than its own. The third condition is if the receiver's queues are full and it cannot handle any more requests. If any of these conditions are met, then the receiver can issue an exception message to the sender to choose an alternate neighbor.

These three conditions in which the receivers will issue exception messages towards their senders will allow each node to make the best decision in terms of choosing a neighbor that is both cost effective distance wise and neighbor conscious in terms of energy and load levels.

## 4.0 EXPERIMENTAL RESULTS

Due to the space constraints of this paper and early stage of our work, our experimental results are very thin from the point of view of comparing our proposed algorithm with related work that has similar goals to ours. We implemented a C++ based simulation with all of the assumptions from section 2.

Table 1 attempts to capture an overview comparison between our diffusion based algorithm and other proposed algorithms, namely Leach [3] and Pegasis [4]. The algorithm names in Table 1 can be briefly explained below.

- Direct [3, 4]: direct communication between each node and the base station
- Leach [3, 4]: clustering-based protocol that utilizes randomized rotation of cluster heads to collect data from neighboring nodes, aggregate data, and send it to the base station
- Pegasis [4]: chain-based protocol, in which group heads are chosen randomly and each node communicates only with a close neighbor forming a chain leading to its cluster head, which in turn communicates with the base station
- MTE [3]: each node sends to the closest neighbor on the way to the base station solely based on spatial information (naïve diffusion)
- Static Clustering [3]: Same as Leach, but with the cluster heads chosen apriori.
- Diffusion: the proposed algorithm in this paper

| Algorithm Name | First Node Dies | Last Node Dies | % of System Lifetime with 100% utility |
|---|---|---|---|
| Direct [4] | 56 | 122 | 45.90% |
| Leach [4] | 690 | 1077 | 64.07% |
| Pegasis [4] | 1346 | 3076 | 43.76% |
| Direct [3] | 217 | 468 | 46.37% |
| MTE [3] | 15 | 843 | 1.78% |
| Static Clustering [3] | 106 | 240 | 44.17% |
| Leach [3] | 1848 | 2608 | 70.86% |
| Diffusion | 3425 | 4323 | 79.23% |

Table 1: Summary view of various algorithms from related work whose goals are similar to ours

The comparison from Table 1 clearly shows the number of iterations the system lived with all nodes being alive and with all nodes being dead. Note that the proposed system lived nearly 80% of its lifetime with 100% utility. The number of iterations each algorithm lasted before the first and last node died should not be compared among the various algorithms since all these results were extracted from various works [3, 4] and this paper. Each paper had their own assumptions about the node characteristics (transmit/receive/amplification power dissipation), and therefore the wide range of system lifetimes. What should be evident is the percentage of time each algorithm allowed the network to operate at 100% utility. The only way to really compare the performance of all these algorithms with our own would be to implement each algorithm using the same assumptions as the ones we used for our algorithm; we will reserve this task for future work.

Figure 2 shows the performance of the system in terms of system lifetime (iterations) and system utility (percentage). We can see that at almost 3500 iterations, the first node dies and in fewer than 900 iterations, the entire system is dead. This is quite impressive in relation to the simplicity of the algorithm and the amount of synchronization messages that are required.

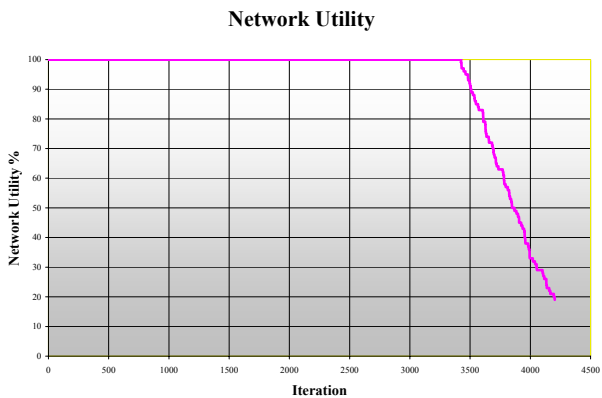

Figure 2: Our enhanced diffusion based routing algorithm system lifetime performance

## 5.0 CONCLUSION AND FUTURE WORK

We have shown a simple and elegant routing algorithm based on local greedy decisions that allows all nodes in the network to deplete their energy supply at roughly the same rate. The algorithm also has little overhead in terms of communication synchronization since most of the synchronization is appended to the actual data transmitted, and therefore it is not very expensive to keep the neighboring nodes synchronized.

As future work, we hope to implement various other routing algorihms, namely the simple direct communication algorithm in which all nodes communicate directly with the base station, and the clustering based routing algorithm similar to LEACH [3] and PEGASIS [4]. Comparing our algorithm with these other routing algorithms will prove to be most valuable and informative. Furthermore, we will address data aggregation and data fusion in the hopes to further reduce the amount of energy that is used in inter-node communication.

## 6.0 REFERENCES

[1] R. Pichna and Q. Wang, "Power Control", The Mobile Communications Handbook. CRC Press, 1996, pp. 370-30.

[2] R. Ramanathan and R. Hain, "Topology Control of Multihop Wireless Networks Using Transmit Power Adjustment", Proceeding Infocom 2000, 2000.

[3] W. Heinzelman, A. Chandrakasan, and H. Balakrishnan, "Energy-Efficient Communication Protocol for Wireless Microsensor Networks", Hawaiian Int'l Conf. on Systems Science, January 2000.

[4] S. Lindsey and C. Raghavendra, "PEGASIS: Power-Efficient GAthering in Sensor Information Systems", International Conference on Communications, 2001.

[5] I. Raicu, O. Richter, L. Schwiebert, S. Zeadally, "Using Wireless Sensor Networks to Narrow the Gap between Low-Level Information and Context-Awareness", ISCA 17th International Conference on Computers and Their Applications, CATA 2002.

[6] J. Hill, R. Szewczyk, A. Woo, S. Hollar, D. Culler, K. Pister, "System architecture directions for network sensors", ASPLOS 2000.

[7] A. S. Tanenbaum, Computer Networks, Third Edition, Prentice Hall Inc., 1996, pp. 276-287.

[8] G Coulouris, J. Dollimore, T. Kindberg, Distributed Systems, Concept and Design, Third Edition, Addison-Wesley Publishers Limited, 2001, pp. 116-119.

[9] C. Intanagonwiwat, R. Govindan and D. Estrin, "Directed diffusion: A scalable and robust communication paradigm for sensor networks", Proceedings of the Sixth Annual International Conference on Mobile Computing and Networking (MobiCOM '00), August 2000, Boston, Massachussetts.